\documentclass[doublecol]{epl2} 
\usepackage{amsmath,amssymb}    
\usepackage[breaklinks,colorlinks = true,linkcolor = blue,urlcolor=blue,citecolor=blue]{hyperref}
\usepackage[capitalize]{cleveref} 

\allowdisplaybreaks
\def \rp   {{\rm p}}
\def \rd   {{\rm d}}
\def \bu   {{\bf u}}
\def \bC   {{\bf C}}
\def \bR   {{\bf R}}

\def \zetap {\zeta_{\rm p}}

\def \taup  {\tau_{\rm p}}

\def \muf  {\mu_{\rm f}}
\def \mup  {\mu_{\rm p}}
\def \mbt  {\frac{\mu_{\rp}}{\tau_{\rp}}}
\def \bx {{\bf x}}
\def \br {\bf r}

\def \bxh {{\bf \hat{x}}}
\def \byh {{\bf \hat{y}}}
\def \bzh {{\bf \hat{z}}}
\def \ABC {\nu[(A\sin z + C\cos y) \bxh + (B\sin x + A\cos z) \byh + (C\sin y + B\cos x) \bzh]}

\def \dissf {\varepsilon_{\rm f}}
\def \dissp {\varepsilon_{\rm p}}
\def \dissphi {\varepsilon_{\phi}}
\def \disst {\varepsilon_{\rm t}}
\def \pass  {\phi}
\def \lap   {\nabla^2}
\def \urms  {u_{\rm rms}}
\def \Rey  {\mbox{Re}}
\def \Deb  {\mbox{De}}
\def \Sc  {\mbox{Sc}}
\def \Rel  {\mbox{Re}_{\lambda}}
\def \dphi {\delta \pass}
\def \drphi {\delta_{\br} \pass}
\def \dphir {\delta_r \pass}
\def \dur {\delta_r u}

\def \flat {\mathcal{F}}

\def \Stwop {S^{\rm P}_{2}}
\def \Stwon {S^{\rm N}_{2}}

\def \Sphip {S^{\phi}_{\rp}}

\def \zetap {\zeta_{\rp}}

\newcommand \adv[1]  {\bu \cdot \nabla {#1}}

\newcommand \norm[1] {\lVert #1 \rVert}
\newcommand \lrp[1] {\left( #1 \right)}
\newcommand \lrs[1] {\left[ #1 \right]}
\newcommand \bra[1] {\left\langle #1 \right\rangle}
\newcommand \pt[1] {\frac{\partial #1}{\partial t}}
\newcommand \DDt[1] {\frac{D #1}{Dt}}

\newcommand \lrv[1] {\left\lvert #1 \right\rvert}
\newcommand \abs[1] {\left\lvert #1 \right\rvert}

\newcommand{\oist}{Complex Fluids and Flows Unit, Okinawa Institute of Science and Technology Graduate University, Okinawa 904-0495, Japan}

\title{Patchy Polymeric Scalar Turbulence} 
\shorttitle{Inefficient Mixing in Polymeric Turbulence at small Sc}
\author{Rahul K. Singh \inst{1} \and Marco E. Rosti \inst{1}}
\shortauthor{R. K. Singh \& M. E. Rosti}
\institute{\inst{1} \oist}
\abstract{
Turbulent polymeric flows show strong deviations from Kolomogorov-like behaviour resulting from more complex dynamics compared to Newtonian turbulence. 
We now study the nature of mixing in polymeric turbulence via Eulerian passive scalar fields of varying molecular diffusivities, given by the Schmidt number Sc. We show that polymeric turbulence is a less efficient mixer than the Newtonian one at small to moderate Sc numbers. Newtonian scalar turbulence (NST) forms large islands of fluctuations with extended, contiguous fronts. In contrast, polymeric scalar turbulence (PST) is marked by small, interspersed patches of strong but less intermittent fluctuations. These patches collectively comprise a larger volume fraction of strong fluctuations, indicating a less efficient mixing, alongwith smaller scalar gradients and therefore smaller average flux across their boundaries. Box counting dimensions reveal a smoother and more space filling nature of patch boundaries in PST compared to NST fronts. Finally, spatial changes of the scalar are stronger in PST, but with a slower self-similar growth and less intermittency as revealed by the kurtosis of scalar differences. Overall, these observations hint at reduced mixing in PST where fluctuations are typically stronger while the average scalar flux is smaller in a stationary state.
}

\begin{document}

\maketitle

Dilute solutions of polymers in Newtonian solvents are well known to exhibit a wide range of intriguing flow phenomena~\cite{Godfrey2008, Benzi2018, Hof2023}. Well known examples range from drag reduction~\cite{Toms1948, Lumley1973, Prasad2006, Bos2025, Casciola2025} at large Reynolds ($\Rey$) numbers to elasto-inertial turbulence (EIT) at moderate $\Rey$~\cite{Samanta2013, Zhang2021, Hidema2024} to elastic turbulence (ET) at vanishingly small $\Rey$~\cite{Steinberg2000, Boffetta2008, Steinberg2021, Soligo2023, RKS2024, Picardo2024, Giulio2024, Giulio2026}. Both ET and EIT are marked by chaotic, mixing flow states that would otherwise remain laminar in the absence of polymers.
A recent addition to this list has been the discovery of an extended, novel self-similarity in polymeric turbulent (PT) flows at very large $\Rey$ via both experiments~\cite{Bodenschatz2021} and simulations~\cite{Marco2023}. This far-from-Kolmogorov behaviour is directly tied to the emergence of a hidden second invariant~\cite{Invariant2025} with an extended Kolmogorov framework capturing the complete elasto-inertial effects~\cite{Ale2025}. 
Elastic effects tend to weaken~\cite{Gillissen21} and slow down~\cite{RKS2025} the inertial, turbulent cascade alongside modifying the topology of turbulence with a quasi-two-dimensionalisation of the small scales of the flow~\cite{Chiarini2025}. 
This is possibly linked to the fact that viscosity dominated small scales in PT in fact manifest ET~\cite{Piyush2025}. 
Surprisingly, small scale intermittency corrections were found to be the same as that in Newtonian turbulence (NT)~\cite{Marco2023}.
Obviously, PT exhibits far more complex and richer dynamics compared to NT, and one expects that such distinctions carry their imprints over all tell-tale attributes of NT. 
We then ask how do these differences alter the \textit{mixing} nature of turbulence. That is, we wish to understand whether or not PT offers any advantageous mixing over NT.

\begin{figure*}[!ht]
	\centering
	\includegraphics[width=1.0\textwidth]{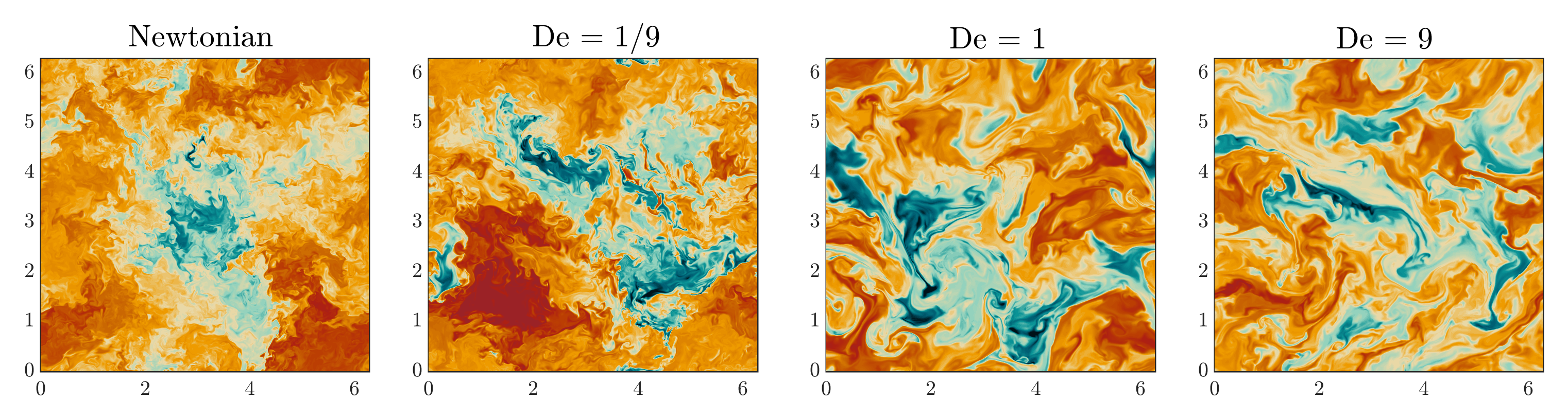}
	\caption{Passive scalar fluctuations about the mean for Sc $= 1$ in polymeric turbulence, with elasticity  increasing from left to right. NST exhibits contiguous fronts separating large islands of similar concentrations, while PST comprises of iso-scalar patches scattered across the domain. The color scale varies linearly in $\lrs{-8,8}$ from deep teal to deep burgundy.}
	\label{fig:Snaps}
\end{figure*}

Efficient mixing is the hallmark of turbulent flows and has been extensively studied~\cite{Ottino1990,Dimotakis2005,Abarzhi2010,Gregg2018,Villermaux2019,Caulfield2021}. 
The simplest and most commonly employed tool to study turbulent mixing is that of a passive scalar which does not couple back to the flow. 
Passive scalar fields often represent diffusive contaminants such as temperature, dye in water, reacting species etc.~\cite{Warhaft2000} and have also helped better our understanding by serving as simplified proxies of NT.
Following the exact results by Obukhov~\cite{Obukhov1949} and Corrsin~\cite{Corrsin1951}, much is now known about passive scalar turbulence (PST), such as anomalous exponents and deviations from the Obukhov-Corrsin theory ~\cite{Falkovich2001, Chertkov2003, Schumacher2010, Iyer2018, Sreei2019, Sreeni2021, Buaria2021}. 
Similar ideas of passive scalar mixing have also been applied to flows of polymeric fluids~\cite{Groisman2004, Teodor2004, Thiffeault2011, Li2017, Picardo2024, Pinho2025, Dzanic2025}.
However, the role of polymers in turbulent mixing at large Re, especially in the regime of PT remains unknown. 
We now answer this question via direct numerical simulations (DNS) of passive scalars advected by a turbulent polymeric flow.  

Polymeric fluid dynamics is governed by the following coupled equations
\begin{align}
		\rho  \DDt \bu  &= - \nabla p + \muf \nabla^2 \bu  + \mbt  \nabla \cdot \bC + \mathbf{F},	 \label{eq:NN}  \\
		\DDt \bC  &= \bC \nabla \bu + (\nabla \bu)^T \bC - \frac{1}{\taup} (\bC -\mathbf{I}), \label{eq:Conf} 
\end{align}
which account for the simultaneous evolution of the incompressible velocity field $\bu (\bx,t)$ (under the constraint $\nabla \cdot \bu = 0$), and the extra stresses due to the polymers given by their conformation tensor $\bC (\bx,t) \equiv C_{ij} = \bra{R_iR_j}$, which is the (configurational) averaged dyadic product of the polymer end-to-end vector $\bR$. 
The average squared end-to-end lengths of the polymers is therefore, $Tr [\bC] = \bra{R_i^2} $, and 
the polymer stress tensor $\frac{\mup}{\taup} (\bC -\mathbf{I})$ corresponds to the Oldroyd-B model of polymers with a relaxation time $\taup$. Note that, the results remain largely independent of the choice of model~\cite{Marco2023,RKS2024}.
We focus on flows with a large value of $\beta \equiv \muf/(\mup+\muf) = 0.9$, that ensures a dilute polymeric solution, where $\mup$ and $\muf$ are the polymeric and fluid viscosities. 
A statistically stationary PT state is sustained via the forcing ${\bf F} = \ABC$, with $A = B = C = 1$. 
The energy is injected at rate $\disst$, and is dissipated away by both the fluid ($\dissf$) and the polymers ($\dissp$), so that $\disst = \dissf + \dissp$. 
Such a turbulent flow is characterised by two dimensionless numbers: the Reynolds number $\Rey,$ which is a measure of turbulence intensity, and the Deborah $\Deb$, that quantifies the polymer relaxation time via $\Deb = \taup/(L/\urms)$, where $L = 2\pi$ is the largest scale of forcing and $\urms$ is the r.m.s. velocity. We fix $\Rel \approx 450$ and $\Deb = 1/9,1,9$ to study the effects of varying elasticity, 
where we have substitutes $\Rey$ by the Taylor-scale Reynolds number $\Rel \equiv \rho \urms \lambda /\muf$, with $\lambda = \urms \sqrt{15 \nu/\dissf}$. 

The PT state thus generated advects a passive scalar field $\pass$, whose time evolution is described by the following forced advection diffusion equation
\begin{align}
	\DDt \phi = \pt{\pass} + \adv {\pass} = \kappa \lap \phi +  f.			\label{eq:PS}
\end{align}
The diffusion constant $\kappa$ defines a third dimensionless number, the Schmidt number Sc $\equiv \nu/\kappa$, defined such that a large value implies a more rough and `turbulent' passive scalar. 
The scalar forcing $f = A \sin(x)$ mimics a large scale $L$ stirring of the scalar. 
We collect the scalar statistics in a stationary state.

\begin{figure}[!ht]
	\centering
	\includegraphics[width=\columnwidth]{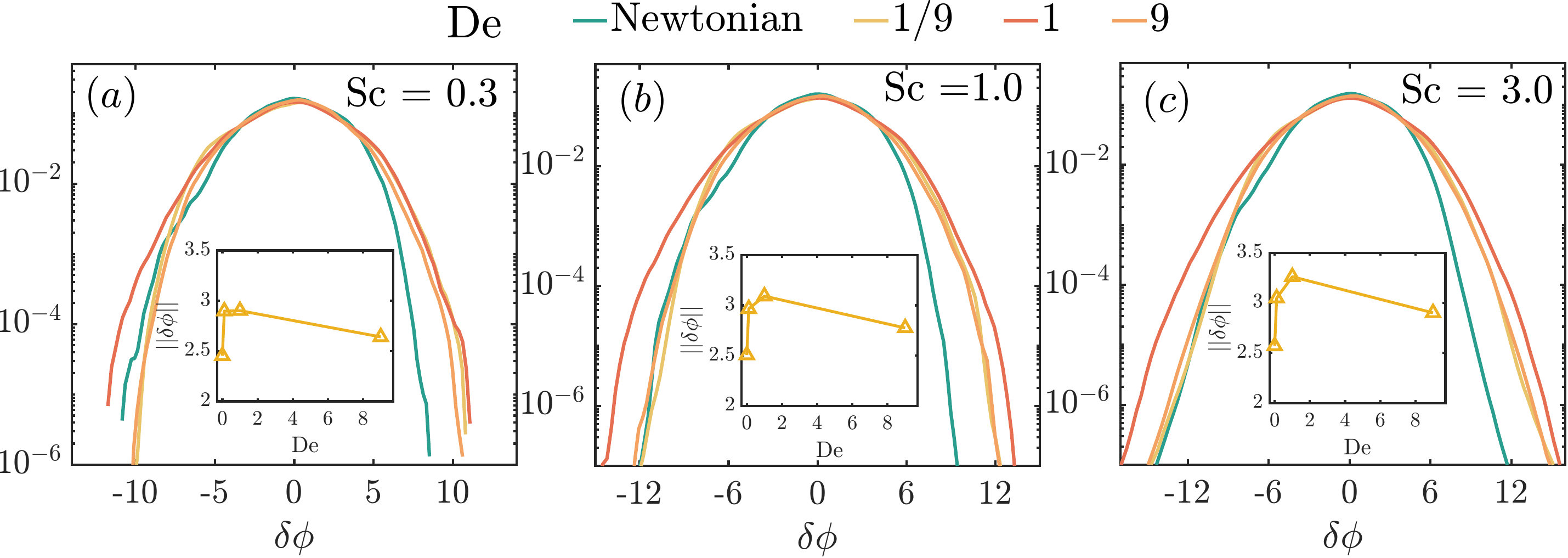}
	\caption{\textbf{Main panels} The probability distribution functions (pdfs) of the passive scalar fluctuations about the mean $\dphi$. A PT background results in stronger fluctuations about the mean, especially at an optimal De $= 1$. \textbf{Insets} The standard deviation $||\delta \phi||$ vs De. }
	\label{fig:Fluc}
\end{figure}
The first glimpses into the nature of polymeric scalar turbulence (PST) is shown in~\cref{fig:Snaps} by the two-dimensional slices of $\phi$ for Sc = 1 where fluid elasticity (De) increases from left to right. 
The colours code the fluctuations about the mean as $\dphi = \pass - \bra{\pass}$, where $\bra{\pass} = 1$. 
The discrepancy between PST and NST (Newtonian scalar turbulence) is evident.
The leftmost panel, corresponding to NST, shows $\dphi$ comprises structures of all scales, ranging from large islands of similar concentration to the very small scale ruggedness/roughness of their contiguous boundaries/fronts. With increasing De, $\phi$-field comprises of smaller, stretched, interspersed patches of stronger fluctuations which are marked by more regular, smoother boundaries. This suggests that mixing in PST is less efficient than NST, since the scalar distribution is less homogeneous. These differences, and more, are quantified in the following sections.
\begin{figure*}[!ht]
	\centering
	\includegraphics[width=\textwidth]{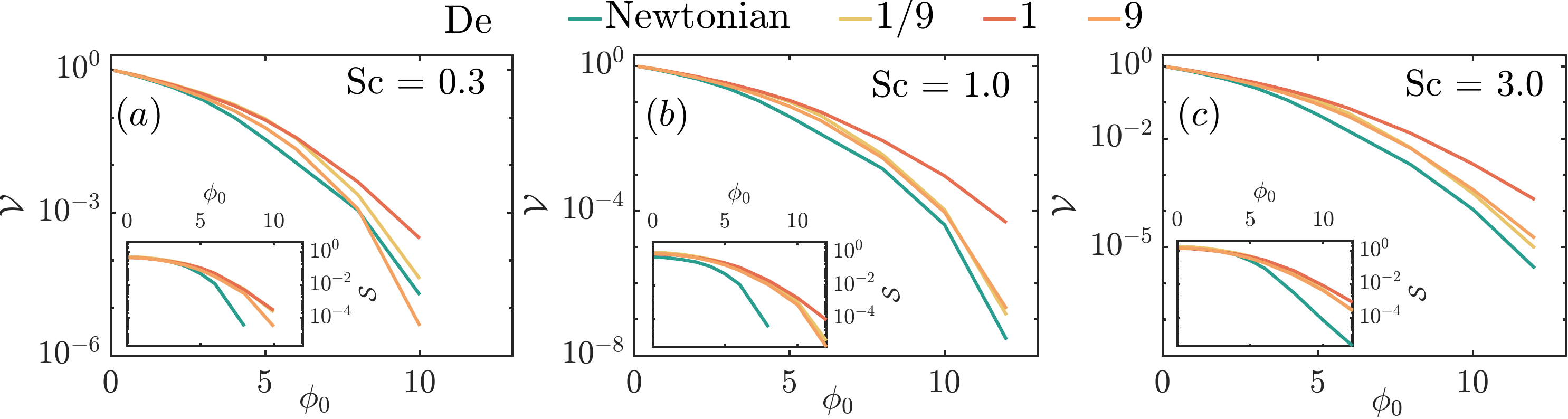}
	\caption{\textbf{Main panels} The fraction of points $\mathcal{S}$ constituting the boundary of regions within which $\delta \phi \geq \phi_0$. \textbf{Insets} Volume fraction $\mathcal{V}$ of all the points in the bulk of the same regions. Larger $\mathcal{S}$ and $ \mathcal{V}$ in PT means regions with large scalar concentration occupy a maximal volume. Thus, mixing is less efficient in PT, and more so at De = 1.}
	\label{fig:AVFrac}
\end{figure*}

\textbf{Fluctuations} 
We quantify the above observations by plotting the probability distribution functions (pdfs) of $\dphi$ for various De and Sc in \cref{fig:Fluc}. 
While all the pdfs peak around $\dphi = 0$, those for PST are visibly wider, with strong fluctuations more probable. 
This means scalars are more likely to concentrate into regions of large concentrations in PST.
The width of the distributions, given by their variance $\norm{\dphi}= \sqrt{\bra{\dphi^2}}$, has a non-monotonic dependence on De, with the pdfs being widest for De = 1. This is shown in the insets of~\cref{fig:Fluc}, where the variance is maximal for $\Deb = 1$ and decreases with De on either side of unity. 
The close to NST statistics at small as well as large De is consistent with the observations in~\cite{Marco2023}, where the fluid energy spectrum falls back to a Newtonian behaviour in these limits. 
As expected, the dependence on Sc is (weakly) monotonic, since the decreasing $\kappa \sim \Sc^{-1}$ is unable to diffuse sharp fluctuations/gradients of the scalar. 
The larger likelihood of strong fluctuations in PST, especially with a maximum variance at De = 1, suggests that mixing is less efficient in PT, as scalars now accumulate more into pockets of large concentrations.

\textbf{Bulk-Boundary Characteristics}
We can verify that there are more regions with large $\abs{\dphi}$ in PST by measuring the volume fraction $\mathcal{V}$ occupied by fluctuations with $\abs{\dphi} \geqslant \phi_0$, for different thresholds $\phi_0 > 0$. 
The volume fraction $\mathcal{V} = n_{bulk}/N^3$, and similarly the fraction of boundary points $\mathcal{S} = n_{bound}/N^3$, can be estimated by identifying all the bulk ($n_{bulk}$) and boundary ($n_{bound}$) points: if any point $\bx$ and all its neighbours satisfy $\abs{\dphi} \geqslant \phi_0$ then it is a bulk point else it is a boundary point~\cite{Schumacher2005}.

We plot $\mathcal{V}$ in~\cref{fig:AVFrac} where the three panels correspond to different Sc and each curve to a different De.
Obviously, $\mathcal{V}$ for large fluctuations increase with Sc as the scalar becomes more turbulent. 
More importantly, extreme fluctuations occupy the smallest volume fraction in NST while it is the largest when De = 1, consistently. 
Then it follows that for small to moderate Sc, NT has a larger mixing efficiency compared to PT.
On either side of De = 1, $\mathcal{V}$ curves again shift closer to the Newtonian. 
Thus, the presence of polymers results in inefficient mixing which is hindered the most when the polymer relaxation time scale is locked with the largest time-scale of the flow.
We also show in the main panels of~\cref{fig:AVFrac} how the bounding surfaces of these volumes have a larger surface area in PST using the count of boundary points $\mathcal{S}$ as a proxy.
This can be attributed to the larger number of isoscalar patches whose common boundaries now occupy a larger cumulative area which is the largest when De = 1.

\begin{figure*}[!ht]
	\centering
	\includegraphics[width=1.0\textwidth]{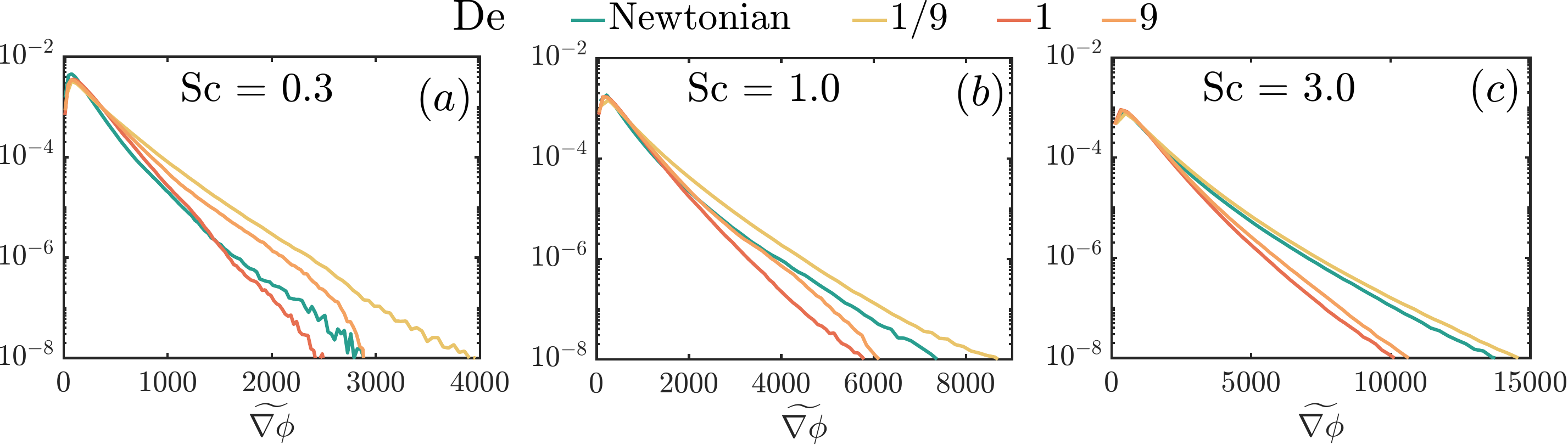}
	\caption{The pdfs of normalised scalar gradients $\widetilde{\nabla \phi (\bx)} = L \lrv{\nabla \phi (\bx)}$ measured at the boundaries enclosing the regions described by $\delta \phi \geqslant 6$ for (a) Sc = 0.3, (b) 1.0, (c) 3.0.}
	\label{fig:Boundgrads}
\end{figure*}
A more direct picture of reduced mixing in PST can be obtained from the rate of transport of the scalar across any point in space. 
Such an understanding of the nature of spatial flux of $\phi$ also has direct implications for mixing, reaction and combustion phenomena~\cite{Pope1988,Picardo2023}. 
We note that the scalar flux across any surface is given as $d\phi (\bx_0) = - \nabla \phi (\bx_0)  \cdot \vec{n}  ds$, where $\bx_0$ is any point on the said surface, and $\vec{n}$ is the normal to an area element $ds$ of the surface. 
We choose the reference boundary surfaces of the regions given by $\dphi \geqslant 4$ where the boundary points are identified in the same as in the preceding paragraph~\cite{Schumacher2005}. 
Scalar gradients, as discussed above, completely determine the statistics of the scalar flux across any given surface. 
We therefore plot the pdfs of the non-dimensional conditioned gradients $\widetilde{\nabla \phi (\bx)} = L \lrv{\nabla \phi (\bx)}_{\bx = \bx_0}$ in~\cref{fig:Boundgrads} across the boundary surfaces enclosing regions where $\dphi \geqslant 6$.
The pdfs show exponential tails at small Sc, that become stretched exponentials for larger values of Sc similarly to what is known for Newtonian flows at small $\Rel$~\cite{Schumacher2005}.
More importantly, flux of the scalar across similar isosurfaces has a maximally wide distributions at small De. 
At larger De, the distributions are less heavy tailed and therefore less intermittent.
That is, the extreme flux events captured by the tails have the least likelihood when De = 1 while small flux regions become more probable.
It is clear that scalar transport is less efficient in PST than NST.

So, mixing is effectively hindered in PST, where scalars form a large number of patches of different concentrations/fluctuations. PST is therefore more likely to store scalars in such patches of strong concentrations with a reduced transport/flux across their boundaries, ultimately reducing the mixing efficiency.

\textbf{Scaling Dimensions}
The boundaries of these scalar patches appear more stretched and less rough in PST in \cref{fig:Snaps}.
One way to compare the roughness is by measuring the scaling dimension $D$ of the number of cubic boxes $N(r)$ of size $r$ that cover an  iso-level set where $N(r) \sim r^{-D}$~\cite{Petre1991,Lohse1994,Iyer2018,Iyer2020}. 
For instance, a space filling and well mixed scalar in three dimensions has $D = 3$ for the isolevels $\dphi = 0$, i.e $N(r) \sim r^{-3}$. A $D < 3$ implies that fluctuations are concentrated into small, intermittent pockets that are spread inhomogeneously~\cite{Iyer2018,Iyer2020} with 
$D = 2 $ implying that isolevels are two-dimensional surfaces, and $1 < D < 2$ loosely means that the surfaces contain holes. 
\begin{figure}[!ht]
	\centering
	\includegraphics[width=\columnwidth]{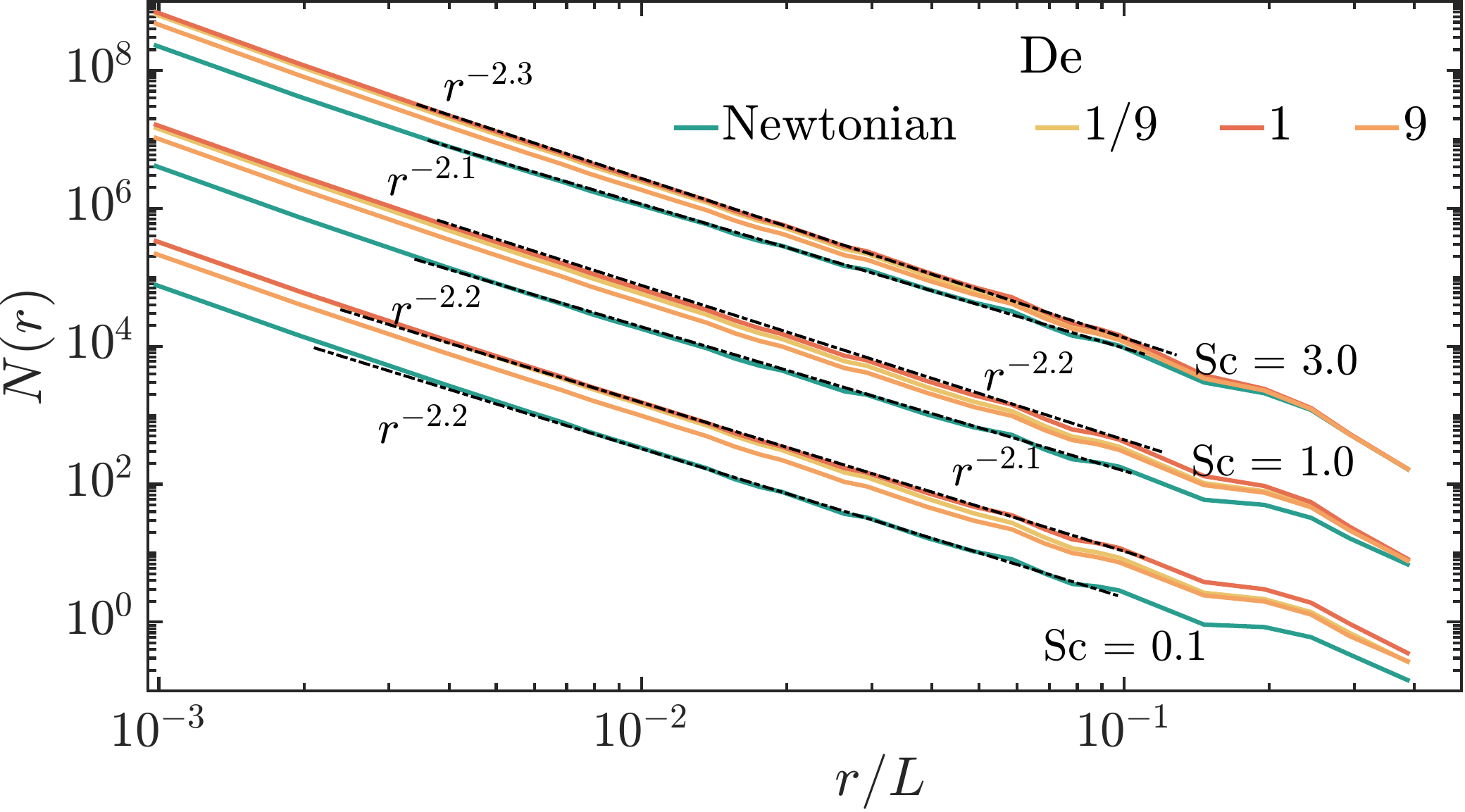}
	\caption{Number $N(r)$ of cubes with side $r$ required to cover the boundaries of regions where $\dphi \geqslant 6$ are maximal at De = 1 for all Sc. A steeper fall-off of $N(r)$ shows patch boundaries are more space filling in PST. The three sets of curves for each Sc have  been moved vertically for visual clarity with only Sc = 1 showing the correct number of boxes.}
	\label{fig:Fractal}
\end{figure}
It is known that $D$ is a function of isolevels themselves, suggesting that there exists an upper bound to mixing in turbulence~\cite{Iyer2020}.
We now plot the number of boxes that cover the above boundary surfaces that enclose the regions $\dphi \geqslant 6$ in~\cref{fig:Fractal}. 
(The curves for Sc = 0.3 and 3.0 have been shifted vertically by factors of 1/20 and 20 respectively.) 
We have checked that boundaries for smaller thresholds do not show much difference across NST and PST and are indistinguishable for $\dphi \lessapprox 4$. 
We see from~\cref{fig:Fractal} that $D \approx 2.2$ remains the same in NST in the range of Sc considered here. 
The PST curves however show a weak Sc dependence, with the curves being the steepest for De = 1. In particular, $D \approx 2.4$ for De = 1 when Sc = 3.0. 
Extreme fluctuations in PST are therefore more space filling, and hence less intermittent. 
These surfaces essentially mark the patch boundaries in PST in~\cref{fig:Snaps} which are visibly smoother and more homogeneously distributed across the domain than the fronts in NST. 
Moreover, at all De and Sc, the PST curves lie above the NST which shows that these isolevels also occupy a larger volume fraction. This observations is consistent with~\cref{fig:AVFrac} where large fluctuations also occupy the largest volume fraction for De = 1 but smallest for NST. 

\textbf{Structure Functions} 
\begin{figure*}[!ht]
	\centering
	\includegraphics[width=0.95\textwidth]{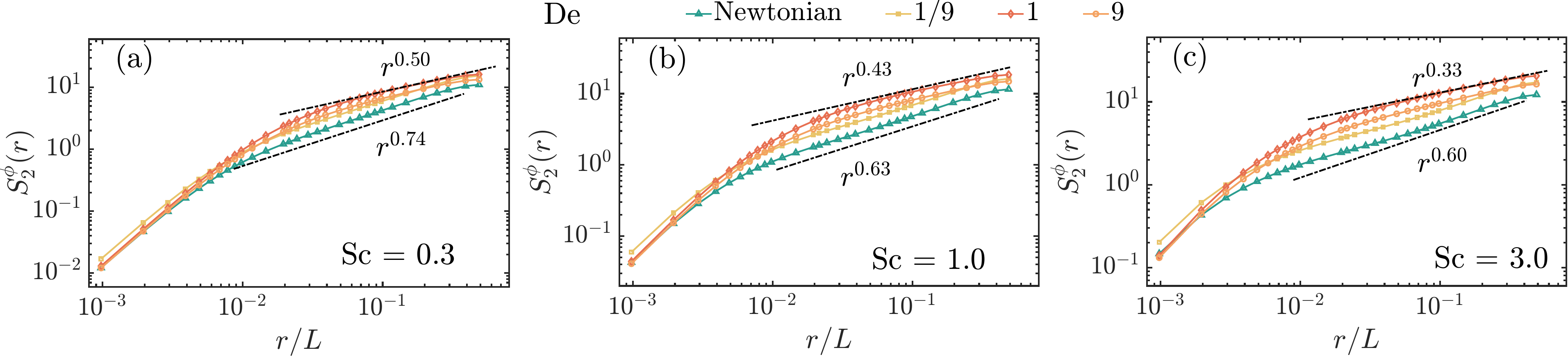}
	\caption{Log-log plots of second-order scalar structure function $\Stwop$ for different De and Sc. The three panels correspond to different Sc numbers. We show the slopes (scaling exponents) via dashed straight lines for De = 1 and Newtonian cases everywhere.}
	\label{fig:Sf}
\end{figure*}
The very different distributions of $\dphi, \nabla \phi$ in PST results a less efficient mixing in PST. 
We now discuss directly how the spatial changes in PST manifestly differ from those in NST.
This can be done via the statistics of scalar difference $\drphi \equiv \phi (\bx+\br) - \phi (\bx)$ which captures the changes in the scalar over a separation $r$. 
The moments of these $r$-increments defined as $\Sphip (r) \equiv \bra{\lrp{\drphi}^{\rp}}$ are referred to as the $\rp$-th order scalar structure functions (SSFs). 
Obukhov~\cite{Obukhov1949} and Corrsin~\cite{Corrsin1951}, using Kolmogorov-like~\cite{K41a} dimensional arguments, asserted that in the inertial-convective range, where both fluid viscosity and the scalar diffusivity have negligible effects, the scalar statistics can only depend on the average rate of scalar (fluctuation) dissipation $\dissphi$ and fluid energy dissipation $\dissf$ in NT.
This means for $L \gg r \gg \eta_{B}$, a simple power counting yields $\Stwon (r) \sim  \dissphi \dissf^{-1/3} r^{2/3}$, where $\eta_{B} = (\nu \kappa^2/\dissf) = \eta_K/\sqrt{\Sc}$ is the Batchelor length scale and $\eta_K = (\nu^3/\dissf)^{1/4}$ is the Kolmogorov length scale. 
It is now well known that NST is strongly intermittent with the exponents deviating significantly from this relation  tendency to saturate~\cite{Iyer2018,Sreeni2021}.
E.g., the observed exponent for $\Stwon \sim r^{\zeta_2}$ is $\zeta_2 \approx 0.6$ which is evidently different from the expected value $2/3$~\cite{Sreeni2021} and is shown in \cref{fig:Sf} by a straight-line guide to the Newtonian curve in the panel (c). We also show plots for Sc = 3 but different De in panel (c) of~\cref{fig:Sf} while panels (a) and (b) report the results for $\Sc = 1/3, 1$ respectively.

The SSFs in PST very clearly depart from and are larger than those in NST with the difference increasing with $r$ and is the largest for De = 1. 
This is supported by~\cref{fig:Snaps} where NST comprises large, continuous islands of iso-colours defined by well defined fronts while the snapshots for De = 1,9 show rather small, isocoloured patches of large fluctuations that are more disconnected and interspersed. 
This means the scalar fluctuations are stronger and happens more often in PST due to larger number of patches.
At small $r$, analyticity requires $S^{{\rm P}/{\rm N}}_2 \sim r^2$ which has the longest range at small Sc (large $\kappa$). 
More importantly, at larger scales $0.04 \lessapprox r/L \lessapprox 0.4$ the SSFs develop a self-similar, power-law regime.
The exponent $\zeta_2$, at least for the largest Sc, is given by the extension of the preceding scaling arguments to PST. 
We recall that velocity fluctuations in PT exhibit a power-law behaviour $\bra{\dur} \sim \gamma^{1/3} r^{2/3}$ where the novel invariant $\gamma$ gives the rate of flux transfer from fluid to the polymer mode~\cite{Invariant2025}. 
PST statistics can therefore be expected to be determined by $\gamma$ and $\dissphi$ as $\Stwop (r) \sim \dissphi \gamma^{-1/3} r^{1/3}$. This is shown for De = 1 in~\cref{fig:Sf}(c) by a dash-dotted line. 
Of course, such a behaviour is expected only for large Sc and deviations are evident at smaller $\Sc$.  
We show the respective power-laws in black, dash-dotted lines s a guide to the eye only for Newtonian ad De = 1 curves which show extended power-law regimes for velocity fluctuations.

\textbf{Scaling Exponents}
\begin{figure}[!ht]
	\centering
	\includegraphics[width= 0.85\columnwidth]{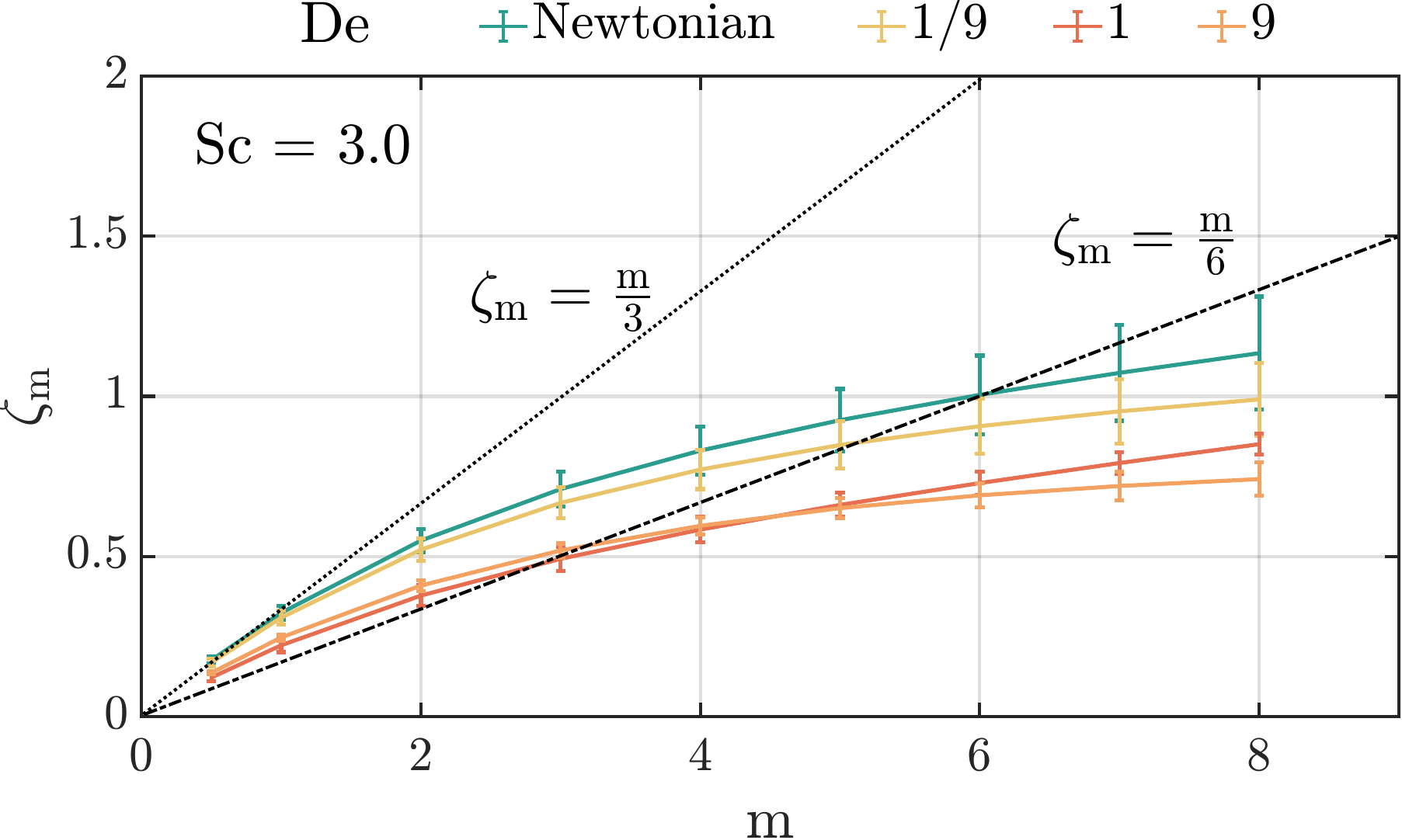}
	\caption{The scaling exponents $\zetap$ for scalar structure functions plotted in~\cref{fig:Sf}. The black dotted line corresponds to Obukhov-Corrsin prediction and its dimensional extension to higher orders $\zetap = \rp/3$ while the dashed line shows the analogous predictions for PST where dimensional arguments yield $\zetap = \rp/6$. }
	\label{fig:zetap}
\end{figure}
To obtain the exact exponents $\zeta_m$, we first compute the local slopes $\tilde{\zeta}_m (r)$ via the log-derivatives $\tilde{\zeta}_m (r) = \rd \log S_m^{{\rm P}/{\rm N}} (r)/\rd \log r$ in the range $\{r/L \ | \ 0.04 \leqslant r/L \leqslant 0.4 \}$.
The mean exponent is given by the average $\zeta_m = \overline{\tilde{\zeta}_m}$ with error bar $\sqrt{\langle (\tilde{\zeta}_m - \zeta_m)^2 \rangle}$. 
The plots of $\zeta_m$ vs $m$ in~\cref{fig:zetap} for  various De and Sc curve away from the dimensional straight line predictions: $\zeta_m = m/3$ for NST and $\zeta_m = m/6$ for PST.
In NST, this results due to the formation and persistence of sharp, intermittent fronts especially at large Sc while a large diffusivity (small Sc) makes the spatial distribution of the scalar more homogenized field by dissipating away those sharp structures~\cite{Warhaft2000,Sreeni1991,Holzer1994,Celani2001,Buaria2021}. 
(We shall return to intermittency in the next section.)
Smaller exponents in PST means that spatial fluctuations grow rather slower in the measured range. For $r_1 < r_2$ in this range
\begin{align*}
\frac{\rd \log \Stwop (r)}{\rd \log r}  &< \, \frac{\rd \log \Stwon (r)}{\rd \log r}  		
\implies	 \frac{\Stwop (r_2)}{\Stwop (r_1)}  < \,   \frac{\Stwon (r_2)}{\Stwon (r_1)}  \\
\frac{\delta \Stwop }{\Stwop (r_1)}    &< \,   \frac{\delta \Stwon }{\Stwon (r_1)}  \ ; \qquad 
\delta \Stwop  = \Stwop (r_2) - \Stwop(r_1)			
\end{align*}
In other words, the relative change in fluctuations is smaller in PST, even though the fluctuations themselves are stronger (cf.~\cref{fig:Sf}). 
This is because of the prevalence of patches in PST within which there is relatively small change in $\phi$. 
So, $\dphir$ between two points changes by a relatively small amount in PST as their separation changes in this range.

\textbf{Flatness}
\begin{figure*}[!ht]
	\centering
	\includegraphics[width=\textwidth]{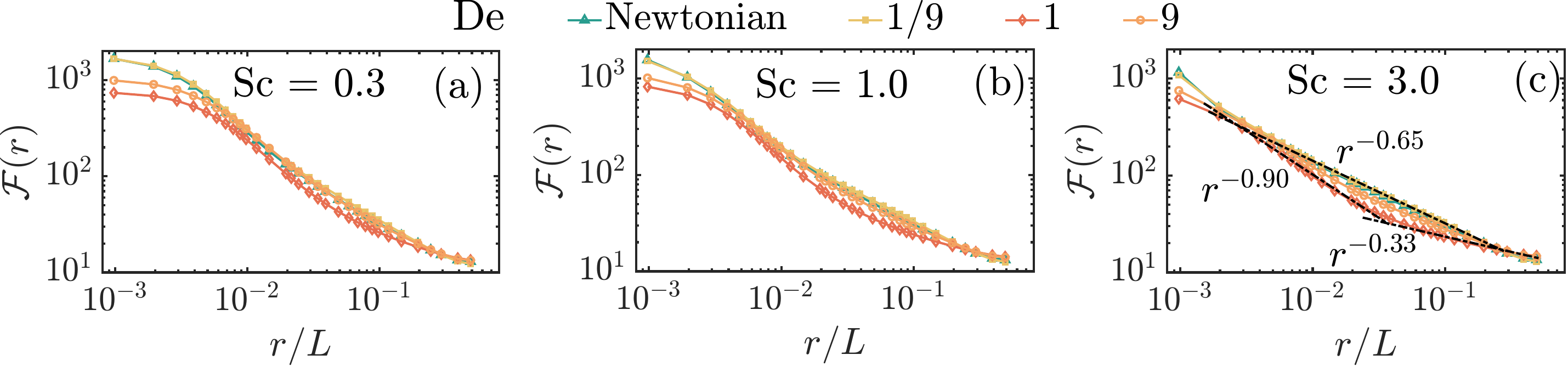}
	\caption{Flatness $\flat (r)$ as a function of $r$ for different De and Sc as the non-dimensionalised ratio of $S^{\phi}_{6} (r)$ and $S^{\phi}_{2} (r)$. Intermittency remains subduedx in PST.}
	\label{fig:ESS}
\end{figure*}
We now ask how extreme and intermittent are the changes in the scalar within these patches in PST.   
In other words, we look at the deviations from the expected power-law behaviours in~\cref{fig:Sf}.
A first answer is given by~\cref{fig:zetap} where the $\zetap$'s in PST visibly deviate lesser from $\zeta_m = m/6$ compared to those in NST whose deviation from $\zeta_m = m/3$ is larger, especially at large Sc. 

That the exponents lie on a curve and not straight lines implies that power-laws are in fact not unique within the patches and the fluctuations scale with a spectrum of exponents in space.
A quantification on how important are these deviations is given by the (kurtosis) flatness $\flat (r)$ of the $\drphi$ distributions $	\flat (r)=\frac{ \bra{\lrp{\dphir}^6}}{\bra{\lrp{\dphir}^2}^3} = \frac{S^{\phi}_{6} (r)}{\lrs{S^{\phi}_{2} (r)}^3} \sim r^{\zeta_6-3\zeta_2}$.
Thus, $\flat (r)$ picks up the relative importance of large spatial changes. 
If $\zeta_m$'s all fall on a straight line, such as $\zeta_m = m/3$, $\flat (r) \sim r^0$. 
This would means fluctuations scale with a unique power-law and are non-intermittent. 
E.g., if the distribution of $\drphi$ is Gaussian at all scales $r$, $\flat (r) = 15$. 
This is observed at very large $r$ where $\flat (r) \approx 15$ irrespective of Sc, De in~\cref{fig:zetap} as changes in the scalar over very large separations are expected to be uncorrelated. 
However, the non-straight-line curves that decrease with $r$ in~\cref{fig:zetap} mean that $\zeta_6  < 3\zeta_2$.
In particular, $\flat (r) \sim r^{\zeta_6 - 3\zeta_2} = r^{-0.65}$ for Sc = 3 in NST for $r/L \gtrapprox  0.01$~\cite{Sreeni2021}.

~\cref{fig:ESS} also shows the $\flat (r)$ plots for different Sc in the three panels; each of which show the curves for different De. 
$\flat (r)$ in PST is bounded by the NST result at all Sc and is minimum for De = 1 everywhere. 
Remarkably, De = 1 is also where PT shows a maximal extent of the novel self-similarity~\cite{Marco2023,RKS2025,Invariant2025}. 
The presence of polymers, therefore, drives fluctuations that are less intermittent. 
Curiously, flatness in PST shows a well developed dual scaling regime for De = 1 and Sc = 3 with $	\flat (r) \sim 	r^{-1/3} $ for $r/L \in \mathbb{S}$ and $\flat (r) \sim 	r^{-0.9} $ for $ r/L \leqslant 0.02$.
Therefore, the growth of $\flat (r)$ (or intermittency) is slower at large scales in PST and much faster for small separations. 
This is because the changes in $\phi$ within and across a patch in PST does not fluctuate strongly which means intermittency is reduced when $r/L \in \mathbb{S}$. 
Now at yet smaller $r$, the extreme changes across boundaries of different patches become important. 
This results in a sharp increase in intermittency for $0 < r/L \lessapprox 0.02$. 
PST, however, is devoid of the long, extended, sharp fronts that separate large islands in NST as can be seen from~\cref{fig:Snaps}.
The fronts are replaced by the boundaries of the stretched isoscalar patches that are more homogeneously distributed and occupy larger volume fractions. 
So, even though $\flat (r)$ grows rather fast in this range, it still remains smaller than that in NST. 
The more homogeneous, and space filling nature of PST could be anticipated from~\cref{fig:AVFrac,fig:Fractal} since large $\lrv{\delta \phi}$ have a larger $\mathcal{V}$ in PST.

In conclusion, we have shown that polymeric scalar turbulence has a very distinct spatial structure compared to the Newtonian counterpart, for when polymeric solutions remain dilute. 
Working with a fixed polymer concentration $\beta = 0.9$, we show that PST comprises of a large number of patches of strong scalar fluctuations that are distinguished by smoother boundaries compared to NST where large islands are separated by very sharp fronts that are rather rough and convoluted. 
Larger number of very different patches in PST means fluctuations, both about the mean and over a finite separation, are larger in PST. 
This is accompanied by a smaller gradient of the scalar in PST across an isoscalar surface which means which the spatial flux of the scalar remains small in PST. 
Thus, mixing remains less efficient in PST where scalars concentrate in patches and their transport is suppressed across patch boundaries.
Finally, we also shed light on the reduced intermittency of spatial fluctuations of the scalar in PST. This can be attributed to the absence of long extended fronts which are replaced by the more space filling patch boundaries in PST. 
Notably, while stronger fluctuations show that large scale mixing of scalars is hindered in PST as they concentrate into interspersed patches, each patch is itself well mixed since the fluctuations grow relatively slow at the scale of patches. 
So, mixing at large scales in PST is hindered, while that at smallest scales is enhanced.
It remains to be seen, however, what is the influence of fluid viscosity and scalar diffusivity in the deep dissipation range, which we barely begin to resolve. It is expected that elastic turbulence takes over in this range~\cite{Piyush2025}. 
As such, scalar statistics in the ET regime in 3D remain yet to be investigated.

\bibliographystyle{eplbib.bst}
\bibliography{refs}

\end{document}